\documentclass{aa}
\usepackage{graphicx}
\usepackage{natbib}
\usepackage{color}
\usepackage{bm,url,times}
\graphicspath{{./fig/}{./png/}}

\newcommand{\EQ}{\begin{equation}}
\newcommand{\EN}{\end{equation}}
\newcommand{\EQA}{\begin{eqnarray}}
\newcommand{\ENA}{\end{eqnarray}}

\newcommand{\Eq}[1]{Eq.~(\ref{#1})}
\newcommand{\Eqs}[2]{Eqs.~(\ref{#1}) and~(\ref{#2})}

\newcommand{\App}[1]{Appendix~\ref{#1}}

\newcommand{\Fig}[1]{Fig.~\ref{#1}}

\newcommand{\Figs}[2]{Figs.~\ref{#1} and \ref{#2}}

\newcommand{\Tab}[1]{Table~\ref{#1}}

{}
{}
{}

{}
{}
{}
{}
{}
{}
{}
{}
{}
{}
{}
{}
{}
{}
{}
{}
{}
{}
{}
{}
{}
{}

{}
{}
{}

{}

{}
{}
{}

\newcommand{\tkapz}{\tilde{\kappa_0}}

\newcommand{\Hminus}{{\rm H}^{-}}
\newcommand{\nnn}{\hat{\mbox{\boldmath $n$}} {}}

\newcommand{\zzz}{\hat{\mbox{\boldmath $z$}} {}}

\newcommand{\xx}{\bm{x}}

\newcommand{\BB}{\bm{B}}

\newcommand{\uu}{\bm{u}}

\newcommand{\JJ}{\mbox{\boldmath $J$} {}}

\newcommand{\AAA}{\mbox{\boldmath $A$} {}}

\newcommand{\FF}{\mbox{\boldmath $F$} {}}

\newcommand{\grav}{\mbox{\boldmath $g$} {}}
\newcommand{\nab}{\mbox{\boldmath $\nabla$} {}}

\newcommand{\SSSS}{\mbox{\boldmath ${\sf S}$} {}}

\newcommand{\DD}{{\rm D} {}}

\newcommand{\dd}{{\rm d} {}}

\def\Hp{H_{\rm p}}

\def\cp{c_{\rm p}}

\def\cs{c_{\rm s}}

\def\rhoe{\rho_{\rm e}}
\def\xHe{x_{\rm He}}
\def\yH{y_{\rm H}}

\def\sigmaSB{\sigma_{\rm SB}}

\def\kB{k_{\rm B}}

\def\chiH{\chi_{\rm H}}

\def\half{{\textstyle{1\over2}}}

\def\onethird{{\textstyle{1\over3}}}

\newcommand{\eV}{\,{\rm eV}}

\newcommand{\kG}{\,{\rm kG}}
\newcommand{\K}{\,{\rm K}}
\newcommand{\g}{\,{\rm g}}
\newcommand{\s}{\,{\rm s}}

\newcommand{\ks}{\,{\rm ks}}
\newcommand{\cm}{\,{\rm cm}}

\newcommand{\km}{\,{\rm km}}

\newcommand{\Mm}{\,{\rm Mm}}

\newcommand{\yapj}[3]{ #1, {ApJ,} {#2}, #3}

\newcommand{\yapjl}[3]{ #1, {ApJ,} {#2}, #3}

\newcommand{\yan}[3]{ #1, {Astron.\ Nachr.,} {#2}, #3}

\newcommand{\yana}[3]{ #1, {A\&A,} {#2}, #3}

\newcommand{\ysph}[3]{ #1, {Solar Phys.,} {#2}, #3}

\newcommand{\yjour}[4]{ #1, {#2}, {#3}, #4}

\newcommand{\ybook}[3]{ #1, {#2} (#3)}

\usepackage{float}

\title{Hydraulic effects in a radiative atmosphere with ionization}
\titlerunning{Hydraulic effects in a radiative atmosphere}
\author{P. Bhat\inst{1,2} \and A. Brandenburg\inst{1,3}}
\authorrunning{Bhat \& Brandenburg}
\institute{
Nordita, KTH Royal Institute of Technology and Stockholm University,
Roslagstullsbacken 23, SE-10691 Stockholm, Sweden
\and
Inter University Centre for Astronomy and Astrophysics,
Post Bag 4, Pune University Campus, Ganeshkhind, Pune 411 007, India
\and
Department of Astronomy, AlbaNova University Center,
Stockholm University, SE-10691 Stockholm, Sweden
}

\date{\today,~ $ $Revision: 1.82 $ $}
\begin{document}

\abstract{
In a paper of 1978, Eugene Parker postulated the need for hydraulic downward
motion to explain magnetic flux concentrations at the solar surface.
A similar process has recently also been seen in simplified
(e.g., isothermal) models of flux concentrations from the
negative effective magnetic pressure instability.
}{
We study the effects of partial ionization near the radiative
surface on the formation of such magnetic flux concentrations.
}{
We first obtain one-dimensional (1D) equilibrium solutions using
either a Kramers-like opacity or the $\Hminus$ opacity.
The resulting atmospheres are then used as initial conditions in
two-dimensional (2D) models
where flows are driven by an imposed gradient force resembling a localized
negative pressure in the form of a blob.
To isolate the effects of partial ionization and radiation, we ignore
turbulence and convection.
}{
In 1D models, due to partial ionization, an unstable stratification forms always
near the surface.
We show that the extrema in the specific entropy profiles
correspond to the extrema in degree of ionization.
In the 2D models without partial ionization, 
strong flux concentrations form just above the height where
the blob is placed. Interestingly, in models 
with partial ionization, such flux concentrations form always
at the surface much above the blob. 
This is due to the corresponding negative gradient in specific entropy.
Owing to the absence of turbulence, the downflows reach transonic speeds.
With $\Hminus$ opacity, flux concentrations are
weaker due to the stably stratified deeper parts.
}{
We demonstrate that, together with density stratification, the imposed
source of negative pressure drives the formation of flux concentrations.
We find that the inclusion of partial ionization affects 
entropy profiles dramatically causing the
strong flux concentrations to form closer to the surface.
We speculate that turbulence effects are needed to limit the
strength of flux concentrations and homogenize
the specific entropy to a more nearly marginal stratification.
}

\maketitle

\section{Introduction}

In a series of a papers, \cite{Par74,Par76,Par78} introduced the idea
of hydraulic flux concentrations at the solar surface.
Here the hydraulic device is formed by magnetic flux tubes of varying
size and pumping is accomplished by turbulence.
In these papers, he envisaged turbulent pumping (analogous to a water
jet vacuum pump) as the relevant driver, but other alternatives such as
the negative effective magnetic pressure instability (NEMPI),
first studied by \cite{KRR89,KMR96}, are possible
and have also been discussed \citep{BGJKR14}.

The flux concentrations of Parker are thought to be just around $100\km$
in diameter.
Such tubes can be concentrated further through what is known as
convective collapse \citep{Par78,Spr79}.
Although these tubes are only about $100\km$, they might be relevant
for sunspots which can be more than a hundred times thicker.
Indeed, in the cluster model of sunspots an assembly of many such smaller
tubes are thought to constitute a full sunspot.
Even today, it is still unclear whether sunspots are monolithic or
clustered \citep[see review by][]{RS11}.
Nevertheless, the possibility of downward flows inside sunspots 
(as seen in Parker's models of hydraulic magnetic flux concentrations) may be a
more universal feature which has also been identified as the driving mechanism
in producing magnetic flux concentrations by NEMPI \citep{BGJKR14} and has 
recently also been seen at the late stages of flux emergence \citep{RC14}.

In most of the work that invokes NEMPI,
an isothermal equation of state is used.
This allows these effects to be studied in isolation from the downdrafts
that occur in convection.
However, it is important to assess the effects of thermodynamics and
radiation, which might either support or hinder tube formation and
amplification.

The goal of the present paper is to investigate how downward flows
produce flux concentrations in a partially ionized atmosphere with
full radiative transfer.
We model the effects of an additional negative pressure by imposing
an irrotational forcing function corresponding
to a localized gradient force of the form $-\nab\phi$
on the right-hand side of the momentum equation, where $\phi$ is a
localized Gaussian profile function that emulates the effects of
negative effective magnetic pressure in a controllable way.
By imposing a vertical magnetic field, we force the resulting flow
to be preferentially along magnetic field lines.
If $\phi$ is chosen to be negative, it corresponds to a negative extra
pressure.
Horizontal pressure balance then leads to a localized gas pressure and
density increase and consequently
to a downflow owing to the weight of this density enhancement.
The return flow closes in the upper parts of this structure.
The resulting flow convergence drives magnetic field lines together
and thus forms the magnetic flux concentration envisaged by
\cite{Par74,Par76,Par78}.
These flux concentrations are also similar to those seen
in studies of NEMPI with a vertical magnetic field \citep{BKR13,BGJKR14}.

We construct hydrostatic equilibrium solutions using a method
similar to that of \cite{BB14}, hereafter BB14.
They fixed the temperature at the bottom boundary, which then also
fixes the source function for the radiation field.
For the opacity we assume here either a generalized Kramers opacity
with exponents that result in a nearly adiabatic stratification
in the deep fully ionized layers.
Alternatively, we use an $\Hminus$ opacity that is estimated from the
number density of $\Hminus$ ions using the Saha equation with a
corresponding ionization potential \citep{KW90}.
For the purpose of our investigation, it is sufficient to restrict
ourselves to the ionization of hydrogen.
This approach was also used by \cite{HNSS07} in simulations of
the fine structure of sunspots.

A general problem in all approaches to time-dependent models of stellar
atmospheres is the large gap between acoustic and thermal timescales.
Their ratio is of the order of the ratio of the energy flux to
$\rho\cs^3$, where $\rho$ is the density and $\cs$ is the sound speed.
For the Sun, this ratio is less than $10^{-10}$ in the deeper parts
of the convection zone \citep{BCNS05}.
This problem has been identified long ago \citep{CS86,CS89} and can be
addressed using models that are initially in equilibrium \citep{NSA09}.
Another possibility is to consider modified models with a larger flux
such that it becomes possible to simulate for a full Kelvin--Helmholtz
timescale \citep{KMCWB13}.
This is also the approach taken here and it allows us to construct models
whose initial state is very far away from the final one, as is the case
with an initially isothermal model.

\section{The model}
\label{RT}

\subsection{Governing equations}

We adopt the hydromagnetic equations for logarithmic density $\ln\rho$,
velocity $\uu$, specific entropy $s$, and magnetic vector potential $\AAA$,
in the form
\EQA
{\DD \ln \rho \over \DD t}&=&-\nab\cdot\uu, \\
\rho{\DD \uu\over \DD t}&=&-\nab(p+\phi) +\rho\grav + \JJ\times\BB
+\nab\cdot(2\rho\nu\SSSS), \\
\label{dss}
\rho T{\DD s \over \DD t}&=&-\nab\cdot\FF_{\rm rad}+2\rho\nu \SSSS^2
+\eta\mu_{\rm M}\JJ^2,\\
{\partial \AAA \over \partial t}&=&\uu\times\BB+\eta\nabla^2\AAA,
\label{sRT}
\ENA
where $\DD / \DD t= \partial /{\partial t} + \nab \cdot \uu$,
$p$ is the gas pressure, $\grav=(0,0,-g)$ is
the gravitational acceleration, $\nu$ is the viscosity,
${\sf S}_{ij}=\half(u_{i,j}+u_{j,i})-\onethird\delta_{ij}\nab\cdot\uu$
is the traceless rate-of-strain tensor and $\SSSS^2={\sf S}_{ij}{\sf S}_{ji}$
contributes to the (positive definite) viscous heating rate,
$\BB=\BB_0+\nab\times\AAA$ is the magnetic field
with $\BB_0=\zzz B_0$ representing an imposed vertical magnetic field,
$\JJ=\nab\times\BB/\mu_{\rm M}$ is the current density,
$\mu_{\rm M}$ is the magnetic vacuum permeability
(not to be confused with the mean molecular weight $\mu$, defined below),
$\eta$ is the magnetic diffusivity,
and $\FF_{\rm rad}$ is the radiative flux.
For the equation of state, we assume a perfect gas with
$p=({\cal R}/\mu)T\rho$, where ${\cal R}=\kB/m_{\rm u}$ is the universal
gas constant in terms of the Boltzmann constant $\kB$ and the atomic mass
unit $m_{\rm u}$, $T$ is the temperature,
and the dimensionless mean molecular weight is given by
\EQ
\mu(\rho,T)=(1+4\xHe)/(1+\yH+\xHe),
\label{mu}
\EN
where $\yH(\rho,T)$ is the ionization fraction of hydrogen and $\xHe$
is the fractional number of neutral helium, which is related to the
mass fraction of neutral helium $Y$ through $4\xHe=Y/(1-Y)$.
In the following, we use the abbreviation $\mu_0=1+4\xHe=(1-Y)^{-1}=X^{-1}$,
where $X$ is the mass fraction of hydrogen (ignoring metals).
In relating various thermodynamic quantities to each other, we introduce
$\alpha=(\partial\ln\rho/\partial\ln p)_T$, which is a known function of $\yH$,
as well as $\delta=(\partial\ln\rho/\partial\ln T)_p$ and the ratio
$\gamma=c_p/c_v$ of specific heats at constant volume and pressure,
$c_v=(\partial e/\partial T)_v$ and $c_p=(\partial e/\partial T)_p$,
respectively, which are known functions of both $\yH$ and $T$;
see \cite{KW90}, \cite{Sti02}, and \App{Thermo}.
When $\yH$ is either $0$ or $1$, we have $\alpha=\delta=1$ and
$c_v=(3/2)\,{\cal R}/\mu$ with $e=c_v T$.
In general, however, we have
$e=(3/2)\,{\cal R}T/\mu+e_{\rm H}$,
where $e_{\rm H}=\yH{\cal R}T_{\rm H}/\mu_0$ is the specific energy
that is used (released) for ionization (recombination)
and $T_{\rm H}=\chi_{\rm H}/\kB$ is the ionization temperature.
Using $\chi_{\rm H}=13.6\eV$ for the ionization energy of hydrogen,
we have $T_{\rm H}\approx1.58\times10^5\K$.

Instead of solving \Eq{dss} for $s$, it is convenient to solve directly
for $T$ using the relation \citep{KW90}
\EQ
\rho T{\DD s\over\DD t}=\rho{\DD e\over\DD t}+p\nab\cdot\uu=\rho c_v T
\left({\DD\ln T\over\DD t}+{\gamma-1\over\delta}\nab\cdot\uu\right).\quad
\label{EntroTemp}
\EN
The pressure gradient is computed as
\EQ
{1\over\rho}\nab p={\cs^2\over\gamma}(\nab\ln\rho+\delta\nab\ln T),
\EN
where $\cs$ is the adiabatic sound speed with $\cs^2=\gamma p/\rho\alpha$.
This approach allows us to find the ionization fraction
of hydrogen from the Saha equation as
\begin{equation}
{\yH^2\over1-\yH}={\rhoe\over\rho}
\left({T_{\rm H}\over T}\right)^{-3/2}
\exp\left(-{T_{\rm H}\over T}\right),
\label{Saha}
\end{equation}
where $\rhoe=\mu_0 m_{\rm u}(m_{\rm e}\chi_{\rm H}/2\pi\hbar^2)^{3/2}$
is the electron density.

To compute $\nab\cdot\FF_{\rm rad}$, we adopt the gray approximation,
ignore scattering, and assume that the source function $S$ is given
by the frequency-integrated Planck function, so $S=(\sigmaSB/\pi)T^4$,
where $\sigmaSB$ is the Stefan--Boltzmann constant.
The negative divergence of the radiative flux is then given by 
\EQ
-\nab\cdot\FF_{\rm rad}=\kappa\rho \oint_{4\pi}(I-S)\,\dd\Omega,
\label{fff}
\EN
where $\kappa$ is the opacity per unit mass
(assumed independent of frequency) and $I(\xx,t,\nnn)$
is the frequency-integrated specific intensity in the direction $\nnn$.
We obtain $I$ by solving the radiative transfer equation,
\EQ
\nnn\cdot\nab I=-\kappa\rho\, (I-S),
\label{RT-eq}
\EN
along a set of rays in different directions $\nnn$ using the
method of long characteristics.
For the opacity, we assume either a Kramers-like opacity
$\kappa=\kappa_0\rho^a T^b$ with adjustable coefficients $\kappa_0$,
$a$, and $b$, or a rescaled $\Hminus$ opacity.
In the former case, following BB14, it is convenient to express
$\kappa$ in the form $\kappa=\tkapz(\rho/\rho_0)^a ({T/T_0})^b$,
where $\tkapz$ is a rescaled opacity and is related to $\kappa_0$ by
$\tilde\kappa_0=\kappa_0\rho_0^a T_0^b$.
With this choice, the units of $\tkapz$ are independent of
$a$ and $b$, and always $\Mm^{-1}\cm^3\g^{-1}$ (=$10^{-8}\cm^2\g^{-1}$).
In the latter case we use for the $\Hminus$ opacity the expression
\citep{KW90}
\begin{equation}
\kappa=\kappa_0\yH(1-\yH){\rho\over\rho_{e^-}}\!
\left({T_{\Hminus}\over T}\right)^{3/2}\!\!
\exp\left({T_{\Hminus}\over T}\right),
\label{Saha2}
\end{equation}
where $\kappa_0=\sigma_{\Hminus}/4\mu_0 m_{\rm u}$ is a coefficient,
$\sigma_{\Hminus}=4\times10^{-17}\cm^2$ is the cross section 
of $\Hminus$ \citep{Mih78},
$x_Z=10^{-4}$ is the fraction of metals,
$T_{\Hminus}=\chi_{\Hminus}/\kB$ and $\chi_{\Hminus}=0.754\eV$ are
the ionization temperature and energy of $\Hminus$,
and $\rho_{e^-}=\mu_0 m_{\rm u}(m_{\rm e}\chi_{\Hminus}/2\pi\hbar^2)^{3/2}$
is the relevant electron density.

An important quantity in a radiative equilibrium model is the radiative
conductivity $K=16\sigmaSB T^3/3\kappa\rho$.
According to the results of BB14, $K$ is nearly constant in the optically
thick part.
This implies that $\rho\propto T^n$ with $n=(3-b)/(1+a)$ being effectively
a polytropic index of the model provided $n>-1$.

For large values of $T$, the exponential terms in \Eqs{Saha}{Saha2}
become unity, and only the terms $1-y_H\propto\rho/T^{3/2}$ from \Eq{Saha}
and an explicit $\rho/T^{3/2}$ term in \Eq{Saha2} remain.
Therefore, $\kappa\propto\rho^2 T^{-3}$, i.e., $a=2$ and $b=-3$,
resulting in a stable stratification with polytropic index $n=(3+3)/(1+2)=2$.

To identify the location of the radiating surface in the model,
we compute the optical depth as
\EQ
\tau(x,z,t)=\int_z^{L_z}(\kappa\rho)(x,z',t)\;\dd z'.
\EN
The $\tau=1$ contour corresponds then to the surface from where most
of the radiation escapes all the way to infinity.
For the forcing function, we assume
\EQ
\phi=\phi_0 e^{-[x^2+(z-z_0)^2]/2R^2},
\EN
where $\phi_0$ is the amplitude with a negative value and $R$ the radius of the blob-like
structure.

\subsection{Boundary conditions}

We consider a two-dimensional (2D) Cartesian slab of size $L_x\times L_z$
with $-L_x/2<x<L_x/2$, $0\leq z\leq L_z$. We assume the 
domain to be periodic in the $x$ direction and bounded by
stress-free conditions in the $z$ direction, so the velocity obeys
\EQ
\partial u_x/\partial z=\partial u_y/\partial z=u_z=0
\quad\mbox{on $\;z=0$, $L_z$}.
\EN
For the magnetic field we adopt the vertical field condition,
\EQ
\partial A_x/\partial z=\partial A_y/\partial z=A_z=0
\quad\mbox{on $\;z=0$, $L_z$}.
\EN
We assume zero incoming intensity at the top, and compute the incoming
intensity at the bottom from a quadratic Taylor expansion of the source
function, which implies that the diffusion approximation is obeyed;
see Appendix~A of \cite{HDNB06} for details.
To ensure steady conditions, we fix temperature at the bottom,
\EQ
T=T_0\quad\mbox{on $z=0$},
\EN
while the temperature at the top is allowed to evolve freely.
There is no boundary condition on the density, but since no mass
is flowing in or out, the volume-averaged density is
automatically constant (see Appendix~C of BB14).
Since most of the mass resides near the bottom, the density there
will not change drastically and will be close to its initial value
at the bottom.

\begin{figure*}[t!]\begin{center}
\includegraphics[width=\textwidth]{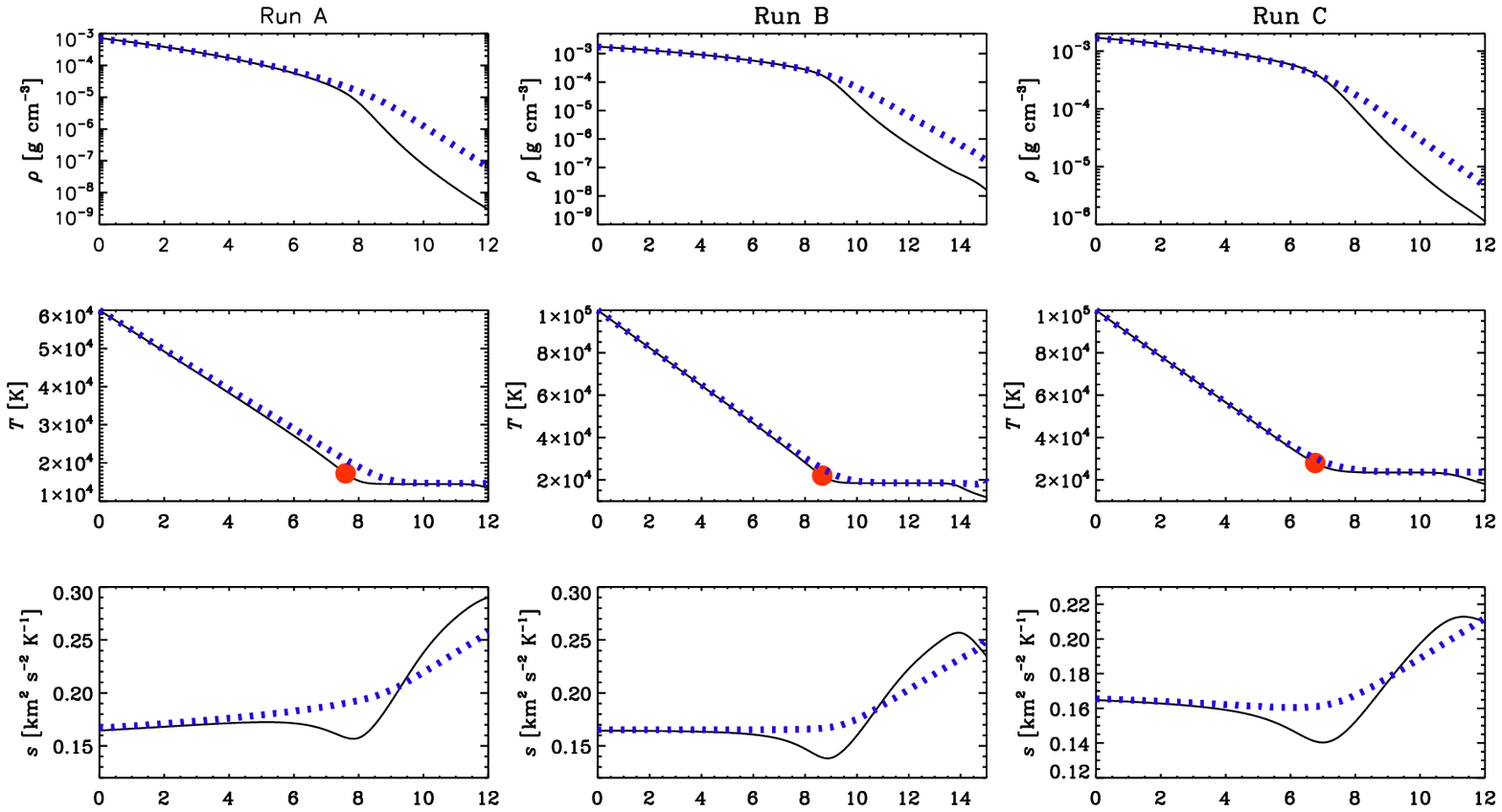}
\includegraphics[width=\textwidth]{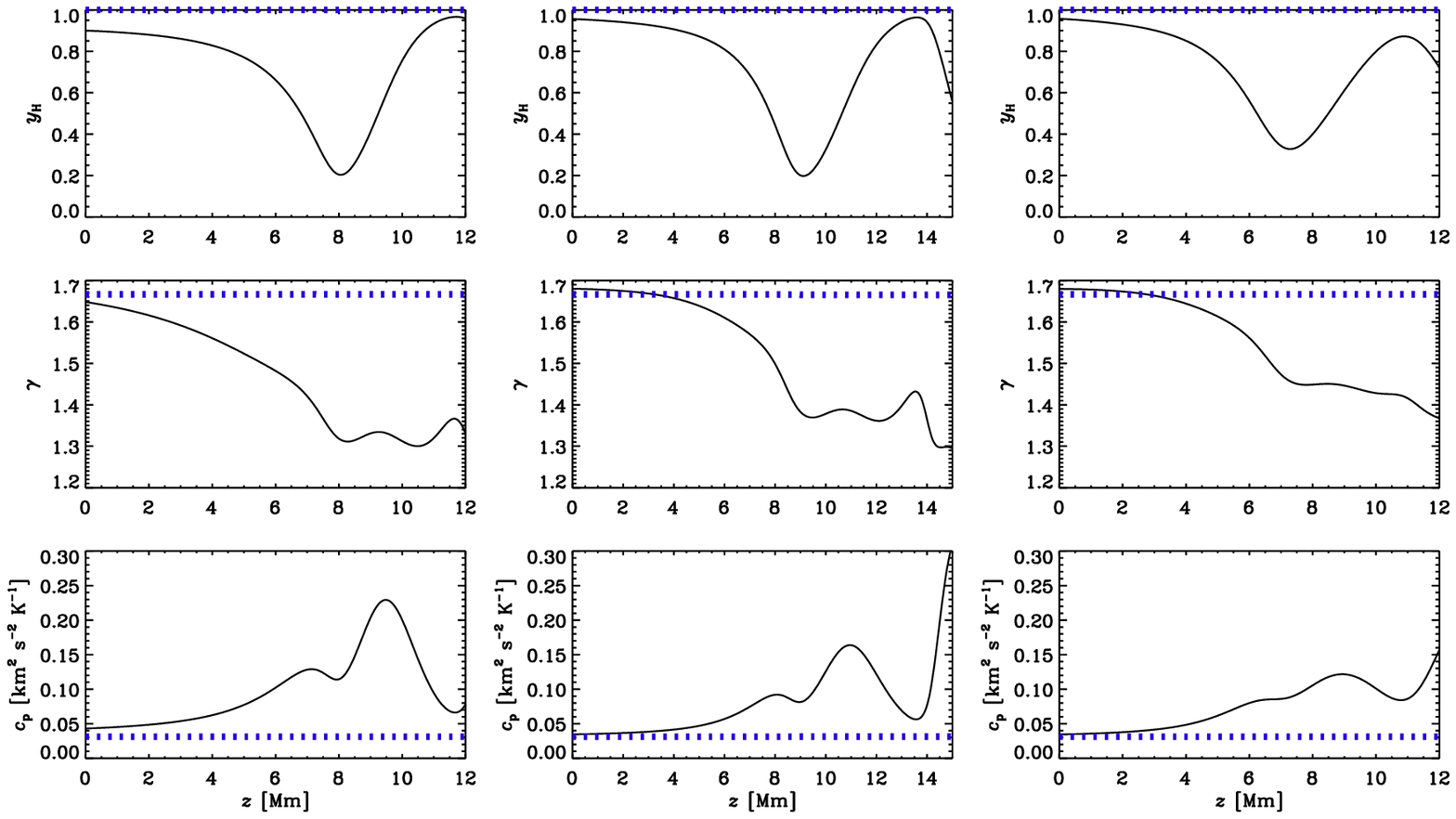}
\end{center}\caption[]{
Comparison of $\rho$, $T$, $s$, $\yH$, $\gamma$ and $c_p$ along each row,
top to bottom, from models with (solid) and without (dotted, blue)
hydrogen ionization for Runs~A, B and C along each column, left to right,
and $\tilde\kappa_0=10^5\Mm^{-1}\cm^3\g^{-1}$.
In the plots of $T(z)$, the closed circles (red) indicate
$\tau=1$.
}\label{profs1_alln}\end{figure*}

We use for all simulations the {\sc Pencil Code}\footnote{
\url{http://pencil-code.googlecode.com/}}, which solves
the hydrodynamic differential equations with a high-order
finite-difference scheme.
The radiation and ionization modules were implemented by \cite{HDNB06}.
All our calculations are carried out either on a one-dimensional (1D)
mesh with $576$ points in the $z$ direction or on a 2D
mesh with $1152\times576$ points in the $x$ and $z$ directions.

\subsection{Parameters and initial conditions}
\label{Parameters}

To avoid very large or very small numbers,
we measure length in $\Mm$, speed in $\km\s^{-1}$,
density in $\g\cm^{-3}$, and temperature in $\K$.
Time is then measured in kiloseconds ($\ks$).
We adopt the solar surface value for the gravitational acceleration,
which is then $g=274\km^2\s^{-2}\Mm^{-1}=2.74\times10^4\cm\s^{-2}$.
In most models we use $(a,b)=(1,0)$ and
$\tilde\kappa_0=10^5\Mm^{-1}\cm^3\g^{-1}$, which yields
a top temperature of about $10,000\K$ (BB14).
We also present results using the $\Hminus$ opacity.
In both cases, the opacities are lowered by 5 to 6 orders of magnitude
relative to their realistic values to
allow thermal relaxation to occur within a few thousand sound travel times.
As discussed in BB14, this also leads to a larger flux and therefore
a larger effective temperature.
For the $\Hminus$ opacity, we have applied a scaling factor
of $10^{-6}$ in \Eq{Saha2}.
In all the models we use $\nu=\eta=10^{-3}\Mm\km\s^{-1}$,
corresponding to $10^{10}\cm^2\s^{-1}$.
For the radius of the blob we take $R=1\Mm$ and for the magnetic field
we take $B_0=1\kG$.

\begin{table}[t!]\caption{
Summary of 1D runs leading to equilibrium solutions.
For the opacity we either give the values $(a,b)$
for Kramers opacity or we indicate $\Hminus$. All runs
are carried out on 576 mesh points. $n$ is the polytropic 
index, $T_0$ and $\rho_0$ are the initial bottom temperature and
density given in $\K$ and $\g\cm^{-3}$, respectively,
and $z_{\tau=1}$ is the height where $\tau=1$.
Runs~A, B and C were carried out with and without ionization.
}\vspace{12pt}\centerline{\begin{tabular}{lcccccc}
Run & opacity & $n$ & $T_0$ & $\rho_0$ & $z_{\tau=1}$ \\
\hline
A & $(1,-7/2)$ & 3.25 & $6\times10^{4}$ & $5\times10^{-4}$ & 7.6 \\ 
B &  (1,0)     & 1.5  & $1\times10^{5}$ & $1\times10^{-3}$ & 8.6 \\ 
C &  (1,1)     & 1    & $1\times10^{5}$ & $2\times10^{-3}$ & 6.8 \\ 
D & $\Hminus$  & ---  & $6\times10^{4}$ & $5\times10^{-4}$ & 7.5 \\ 
E & $\Hminus$  & ---  & $1\times10^{5}$ & $2\times10^{-3}$ & 13.8 \\ 
\label{T1summary}\end{tabular}}\end{table}

\begin{table}[t!]\caption{
Summary of 2D models discussed in this paper.
For the opacity we either give the values $(a,b)$
for Kramers opacity or we indicate $\Hminus$.
$\phi_0$ and $p(z_0)$ are given in units of $\g\cm^{-3}\km^2\s^{-2}$,
while $\max|u_z|$ is given in $\km\s^{-1}$.
}\vspace{12pt}\centerline{\begin{tabular}{lccccccc}
Run & opacity & $z_0$ & $-\phi_0$ & $p(z_0)$ & $\max|u_z|$ \\
\hline
F3  &  (1,0)  & 3 & $3\times10^{-3}$ &1.00& 1.1\\ 
K3a &  (1,0)  & 3 & $3\times10^{-3}$ &0.94& 45 \\ 
K3b &  (1,0)  & 3 & $3\times10^{-4}$ &0.98& 40 \\ 
H3  &$\Hminus$& 3 & $7\times10^{-4}$&0.06 & 22 \\ 
H10 &$\Hminus$&10 & $3\times10^{-2}$&0.07 & 15 \\ 
\label{T2summary}\end{tabular}}\end{table}

\section{Results}

\begin{figure*}[t]\begin{center}
\includegraphics[width=\textwidth]{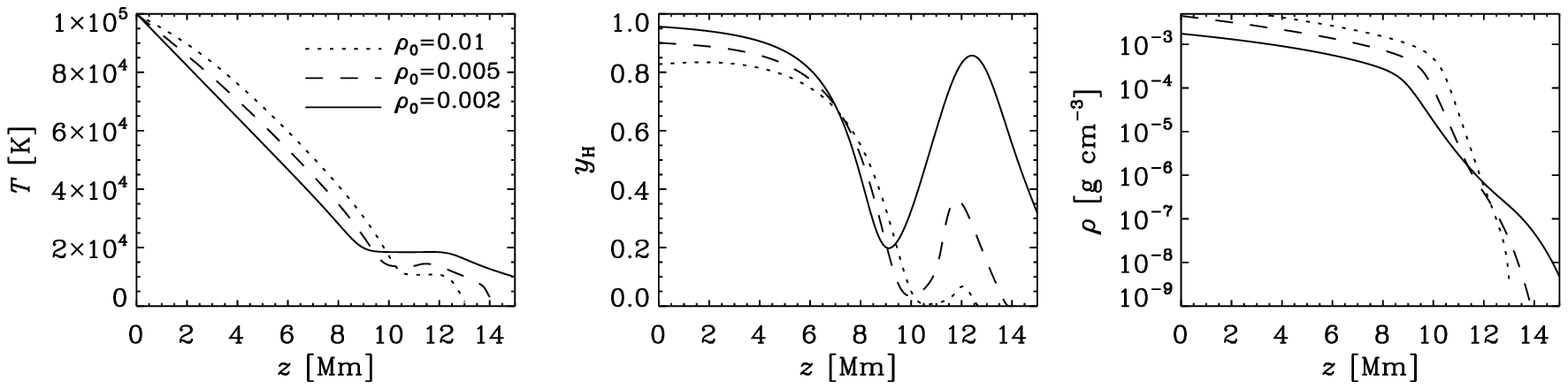}
\end{center}\caption[]{
Comparison of models with $\rho_0=0.002\g\cm^{-3}$ (solid),
$0.005\g\cm^{-3}$ (dashed), and $0.01\g\cm^{-3}$ (dotted lines).
Here $(a,b)$ is $(1,0)$, corresponding to $n=1.5$.
}\label{pcomp_rho}\end{figure*}

First we run 1D simulations with $\phi_0=0$ and isothermal
initial conditions using Kramers opacity and $\Hminus$ opacity. 
A summary of these runs is listed in \Tab{T1summary}.
We use the resulting equilibrium solutions from the 1D runs
as initial conditions for the 2D runs with $\phi_0\neq0$.
The summary of 2D runs is listed in \Tab{T2summary}.

\subsection{Kramers opacity}

As listed in \Tab{T1summary}, for Kramers opacity we use
three pairs of $(a,b)$, $(1,-3.5)$, $(1,0)$, and $(1,1)$
in Runs~A, B, and C, respectively.
In the absence of ionization, the resulting equilibrium solutions
have an optically thick part that is nearly polytropically stratified,
i.e., $\rho\propto T^n$, where $n=(3-b)/(1+a)=$3.25, 1.5 and 1,
respectively are the polytropic indices (BB14) for Runs~A, B and C.
In the outer, optically thin part, the temperature in all cases is nearly
constant and approximately equal to the effective temperature.
For $\gamma=5/3$, a polytropic index of $3/2$ corresponds to
an adiabatic stratification.
The pressure scale height, $\Hp={\cal R}T/\mu g$, in the case of
$n=1.5$ is about $3\Mm$ in the upper parts of the model 
and increases to about $7\Mm$ at the bottom.
In the 2D runs, for the Kramers opacity, we have used only
$(a,b)$ of $(1,0)$, corresponding to $n=1.5$.

In \Tab{T1summary} we also list the height $z_{\tau=1}$ of the
photosphere, where $\tau=1$.
For our models with Kramers opacity, the value of $z_{\tau=1}$
is around $8\Mm$, but comparing the models with $T_0=10^5\K$
and $\rho_0=0.002\g\cm^{-3}$ 
using either Kramers or $\Hminus$ opacity (Runs~C or E, respectively), we find
that $z_{\tau=1}$ doubles from about $7\Mm$ to $14\Mm$,
which is the reason we will choose
a shallower domain for our 2D experiment.

\subsubsection{Vertical equilibrium profiles}

In \Fig{profs1_alln} we compare vertical profiles of various
thermodynamic parameters in 1D models with (in solid black)
and without (in dotted blue) partial ionization with $\phi_0=0$.
Both models have in common that the temperature decreases approximately
linearly with increasing $z$ and then reaches a constant at a height where 
$\tau=1$ (in the one with ionization);
this height is nearly the same in both cases.
By requiring thermostatic equilibrium,
\Eq{dss} yields $\nab\cdot\FF_{\rm rad}=0$,
and in the absence of ionization,
it is seen that the solutions for the temperature profiles
are linearly decreasing for $\tau \gg 1$ and nearly constant for $\tau \ll 1$ (BB14).
The inclusion of ionization does not seem to affect the solutions for
temperature profiles much.
It can be seen that the polytropic density-temperature relation, $\rho\sim T^n$,
nearly follows in the optically thick part ($\tau > 1$) across all
atmospheres with different polytropic indices.
This is because in the optically thick part, the degree of ionization,
$\yH$ remains nearly constant.

In the optically thin part, the models with ionization have lower densities
compared to the models without ionization, thus increasing the density contrast.
The specific entropy in the optically thick part is stratified according to
the respective polytropic indices (stable when $n=3.25$,
marginal when $n=1.5$, and unstable when $n=1$; cf.\ BB14).

Interestingly with ionization, all the entropy profiles in
Runs~A, B, and C behave in a similar fashion
near and above the height where $\tau=1$.
Near $\tau=1$, there is a narrow layer where the vertical entropy gradient
is negative, corresponding to Schwarzschild-unstable
stratification and the possibility of convection.
(We confirmed this and will comment on it in the discussion.)
It can be seen from \Fig{profs1_alln}, that on comparing the
specific entropy profiles with the $\yH$ profiles,
the extrema in the entropy profiles 
coincide with the ones in the corresponding $\yH$ profiles.
This correspondence in the extrema between the two quantities,
specific entropy and degree of ionization can be shown mathematically.
We show in detail in \App{ionssprof} that,
using the equation of state, the first law of thermodynamics, and
the Saha ionization equation, for the case of $\tau\gg1$,
\EQA
\dd s=\dd\yH{{\cal R}\over\mu_0}\left[{1\over A_v}{(n-3/2)\over(n+3/2)}
- B_v\right],
\label{dsdh1}
\ENA
where $A_v$ and $B_v$ are coefficients that are defined
in \Eq{AB} of \App{Thermo}.
In the case of $\tau \ll 1$, we have
\EQ
\dd s=\dd\yH{{\cal R}\over\mu_0}\left[{1\over A_v}- B_v\right].
\label{dsdh2}
\EN
From \Eqs{dsdh1}{dsdh2}, we find that the change in specific
entropy is directly proportional to the change in degree of
ionization and when $\dd \yH=0$, then $\dd s=0$. Thus, the extrema
in $s$ directly correspond to extrema in $\yH$.
The hydrogen ionization fraction $\yH$, reaches a minimum of about 0.2 
(in Runs~A and B) and about 0.4 (Run~C) near
the surface, but then increases again.
This is because of a low density and the exponential decrease in the
upper isothermal layer, leading to larger values of $\yH$ even when
$T$ is small.
In the Sun, the surface temperatures are of course smaller still,
and therefore $\yH\approx0$ can then be reached.
While the specific heats increase outward by a factor of about 5 to 10,
their ratio, $\gamma$, decreases below the critical value of 5/3.

In \Fig{pcomp_rho} we compare models with three values of $\rho_0$.
We recall that $\rho_0$ is the bottom value of the
density of the initially isothermal model.
Since temperature $T_0$ is fixed at the bottom, the pressure scale
height remains unchanged, but since the stratification evolves to
a nearly adiabatic one, the density scale height becomes larger than
the pressure scale height, so density drops more slowly and the
bottom density becomes smaller by about 2/3; a corresponding expression
for this is given by Eq.~(C.5) in BB14.
Note that models with larger values of $\rho_0$ result in lower surface
temperatures and lower degrees of ionization near the surface.
However, for a given number of mesh
points the height of the computational
domain has to be reduced for larger values of $\rho_0$, because the
density drops now much faster to small values.
This is just a numerical constraint that can be alleviated by using
more mesh points.

\begin{figure}[t!]\begin{center}
\includegraphics[width=\columnwidth]{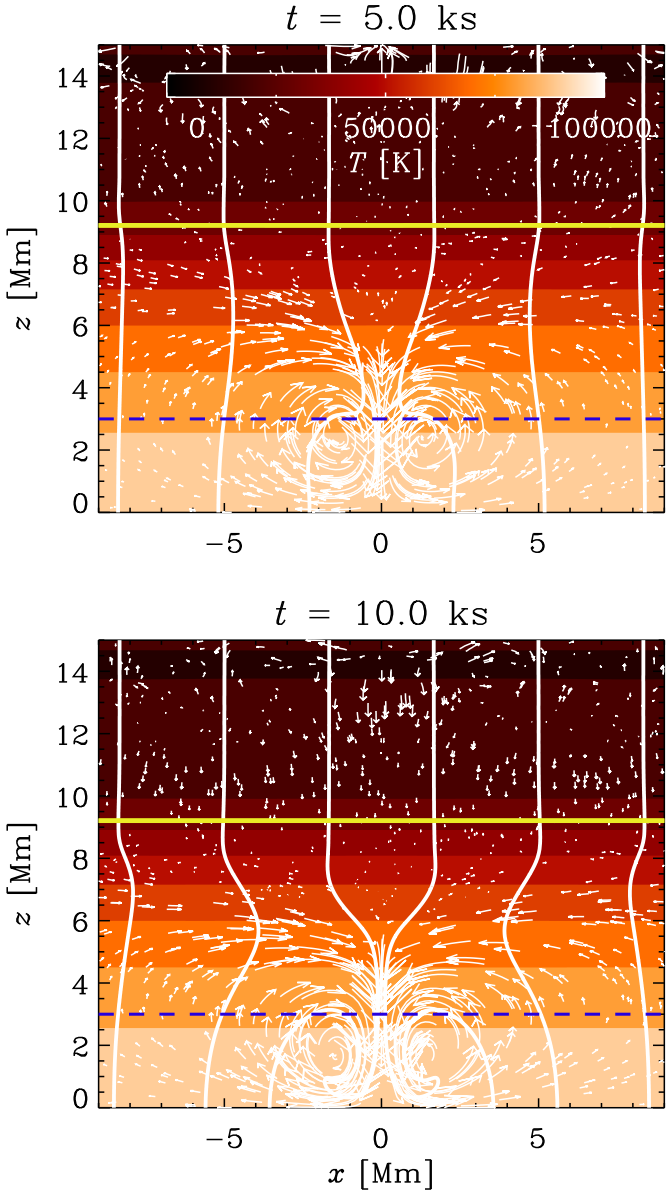}
\end{center}\caption[]{
Snapshots of Run~F3 (fixed ionization, $\yH=1$)
showing temperature (color coded), magnetic field
lines, and velocity vectors at two times before and after the flux
concentration develops.
The solid yellow line at $z\approx9\Mm$ indicates the $\tau=1$ surface while
the dashed blue line indicates the height $z_0$ where suction operates.
}\label{pplane_yHfix}\end{figure}

\subsubsection{Two-dimensional models}

\begin{figure}[t!]\begin{center}
\includegraphics[width=\columnwidth]{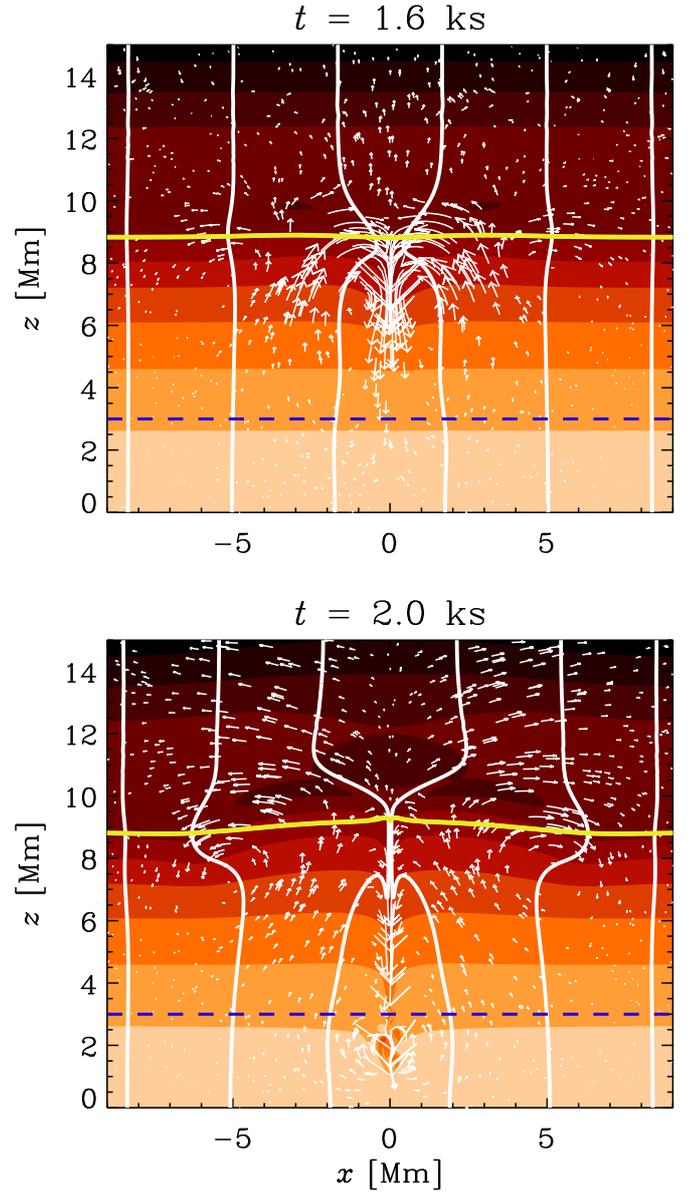}
\end{center}\caption[]{
Same as \Fig{pplane_yHfix}, but for variable partial ionization for
Run~K3a at two times just before and after the flux concentration develops.
}\label{pplane}\end{figure}

Next, we consider 2D models with $\phi_0\neq0$.
The 1D vertical equilibrium solutions form the initial condition
here along $z$ for all $x$.
We consider first the case $\phi_0=-3\times10^{-3}\g\cm^{-3}\km^2\s^{-2}$
using $z_0=3\Mm$ for the height of the blob.
In \Fig{pplane_yHfix} we show the result for Run~F3 (fixed ionization,
$\yH=1$) at $t=5\ks$ and $10\ks$, while in \Fig{pplane} we show the result
for Run~K3a with partial ionization effects included at $t=1.6\ks$ and $2\ks$.
 
In both the cases (runs F3 and K3a in \Figs{pplane_yHfix}{pplane}),
we see the effects of downward suction.
We also see how the magnetic field lines are being pushed together
at a place above the blob where the return flow tries to replenish
the gas in the evacuated upper parts.
In the case of partial ionization (Run~K3a in \Fig{pplane}),
the upper parts have a strongly
negative specific entropy gradient leading to an effect that is
most pronounced at a height considerably above the height of the blob.
Thus, as compared to the case without partial ionization (Run~F3), 
the inclusion of partial ionization (Run~K3a) causes the
flux concentrations to form at the $\tau=1$ surface.
In the Run~K3a at the later time, however, when the magnetic structure
has collapsed almost entirely, the converging inflow has stopped and
there are now indications of an outflow.

\begin{figure}[h!]\begin{center}
\includegraphics[width=\columnwidth]{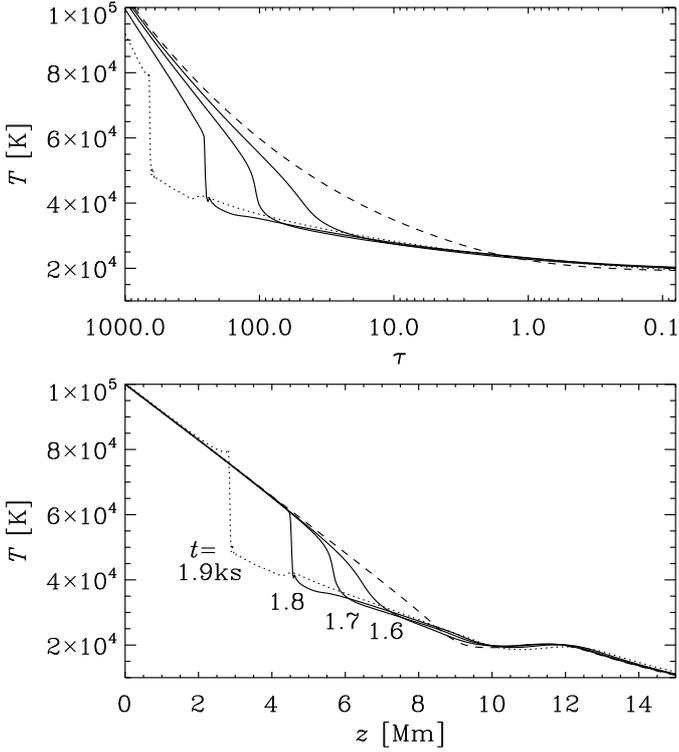}
\end{center}\caption[]{
Temperature versus optical depth and height at different times through
$x=0$ (solid lines and a dotted line for the last time) and $x=L_x/2$
(dashed lines) for Run~K3a at times $1.6$--$1.9\ks$.
}\label{pprofs_TT}\end{figure}

\begin{figure}[h!]\begin{center}
\includegraphics[width=\columnwidth]{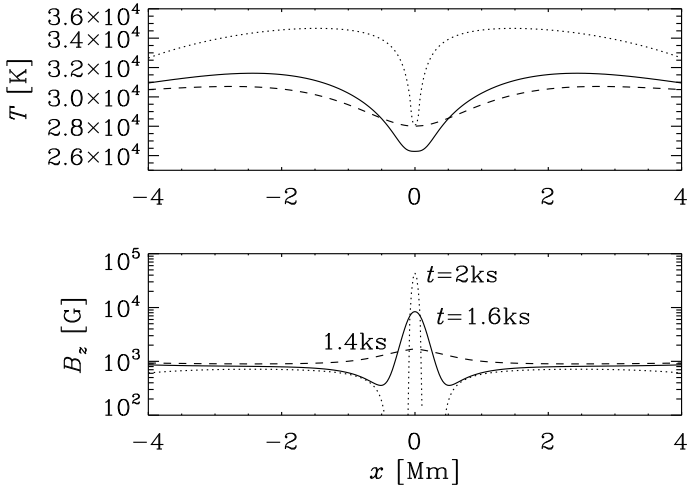}
\end{center}\caption[]{
Temperature and vertical magnetic field strength versus $x$
at different times ($1.4$--$2\ks$) through $z=8\Mm$ for Run~K3a.
}\label{pxprofs_TT_Bz}\end{figure}

\begin{figure}[h!]\begin{center}
\includegraphics[width=\columnwidth]{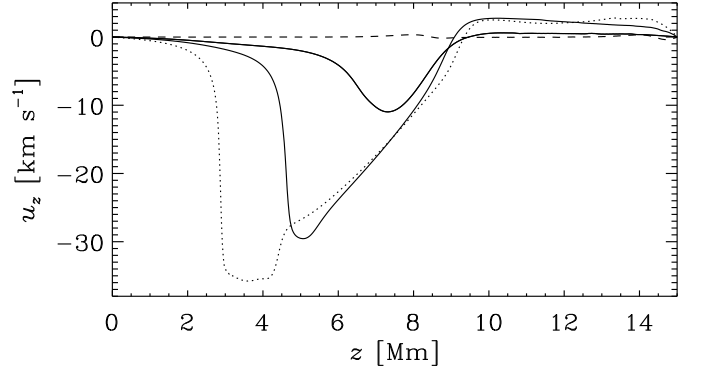}
\end{center}\caption[]{
Vertical velocity versus height at different times ($1.6$--$1.9\ks$)
for Run K3a.
}\label{pprofs_uz}\end{figure}

\begin{figure}[h!]\begin{center}
\includegraphics[width=\columnwidth]{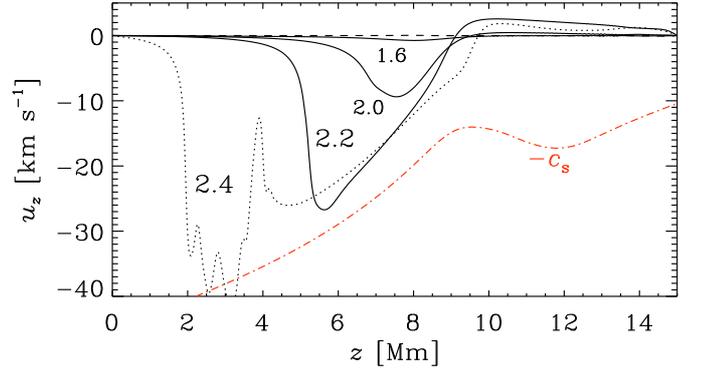}
\end{center}\caption[]{
Similar to \Fig{pprofs_uz}, but for Run~K3b with 10 times weaker suction.
The red dash-dotted lines shows the profile of sound speed.
}\label{pprofs_uz_3em4}\end{figure}

\begin{figure}[h!]\begin{center}
\includegraphics[width=\columnwidth]{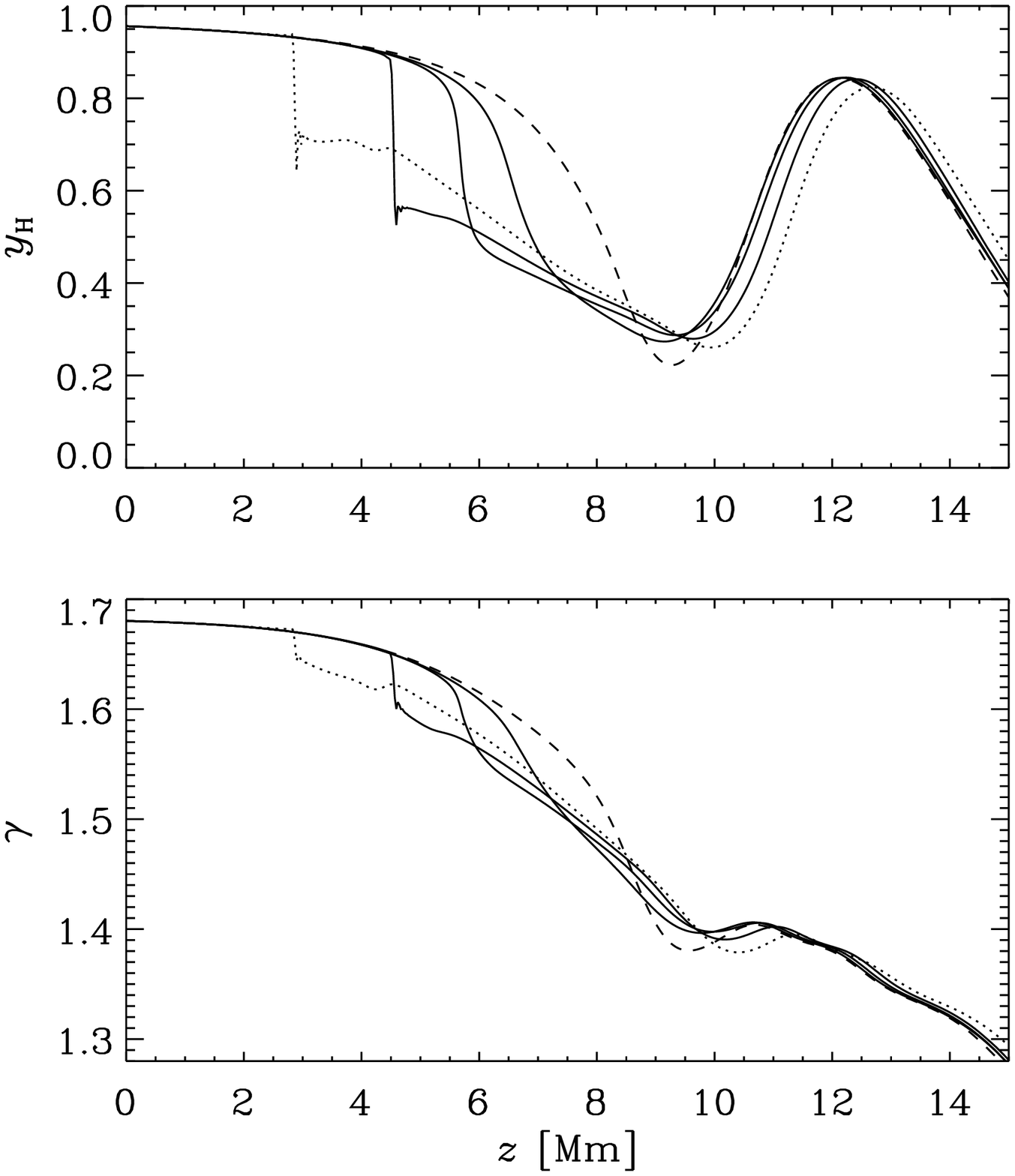}
\end{center}\caption[]{
Degree of ionization and ratio or specific heats versus height, from Run K3a.
}\label{pprofs_yH}\end{figure}

It is remarkable that at all times, the $\tau=1$ surface is approximately
flat, so there is no Wilson depression in our models.
To examine whether this is an artifact of the rather small values of
opacity in our models, which results in comparatively
larger radiative flux and radiative diffusivity,
and therefore horizontal temperature equilibration, we ran a similar
model, using however only vertical rays in the solution of \Eq{RT-eq}.
However, the results were virtually identical, suggesting that the
absence of Wilson depression is not connected with the enhanced luminosity
of our models that is used to reduce the Kelvin--Helmholtz timescale.

In \Fig{pprofs_TT} we show for Run~K3a vertical temperature profiles
though $x=0$ (i.e., through the structure) and $x=L_x/2$ (away from it)
as functions of $\tau$ and $z$.
At $x=0$, we clearly see that for $\tau\gg1$, the temperature drops
progressively below the value at $x=L_x/2$.
At $z=z_0$, the temperature is below $50,000\K$ at $x=0$,
while at $x=L_x/2$ we have $80,000\K$.
Note also that for $\tau<1$, the temperature is slightly enhanced
at $x=0$ compared to $x=L_x/2$.
This is expected, because here the vertical gradient of specific
entropy is positive, corresponding to stable stratification,
so any downward motion would lead to enhanced entropy and temperature
at that position.

In \Fig{pxprofs_TT_Bz} we show the corresponding temperature and magnetic
field profiles through a horizontal cut at $z=8\Mm$, which is just
beneath the surface.
Note that the temperature is reduced at the location of the structure,
but there is also an overall increase in the broader surroundings of the
structure, which we associate with the return flow from deeper down.
The magnetic field enhancement reaches values of the order of about
$50\kG$ (an amplification by a factor of 50) in a narrow spike.
These structures are confined by the strong converging return flow.

The downward speed can become comparable with the local sound speed;
see \Figs{pprofs_uz}{pprofs_uz_3em4}, where we compare two cases
with different forcing amplitudes.
Nevertheless, in both cases the speeds are similar.
This implies that the vertical motion is essentially in
free fall.
To verify this, we note that the speed of a body freely falling over
a distance $\Delta z$ is $v_{\rm ff}=\sqrt{2g\Delta z}$.
Using $\Delta z=5\Mm$, we find $v_{\rm ff}\approx50\km\s^{-1}$,
which is comparable with the speeds seen in \Figs{pprofs_uz}{pprofs_uz_3em4}.
As expected from earlier polytropic convection models with ionization
\citep{RT93}, the downflow advects less ionized material of lower
$\gamma$ and larger $\cp$ downward; see \Figs{pprofs_yH}{pprofs_ss}.
Then again from time evolution plots of $s$ and $\yH$ shown in
\Figs{pprofs_yH}{pprofs_ss}, we find a correspondence between the profiles of
specific entropy and $\yH$, as expected according to \Eqs{dsdh1}{dsdh2}.
Not surprisingly, the suction-induced downflow leads to values of $s$ that,
at larger depths inside the structure, agree with the photospheric values
higher up.
However, temporal changes in $\gamma$ are not as dramatic as the changes
with height.
Inside the structure, the specific entropy has photospheric values also
deeper down, and $s$ is nearly constant (about $0.14\km^2\s^{-2}\K^{-1}$)
in the range $3\Mm\leq z\leq9\Mm$ at $t=1.9\ks$.

\begin{figure}[t!]\begin{center}
\includegraphics[width=\columnwidth]{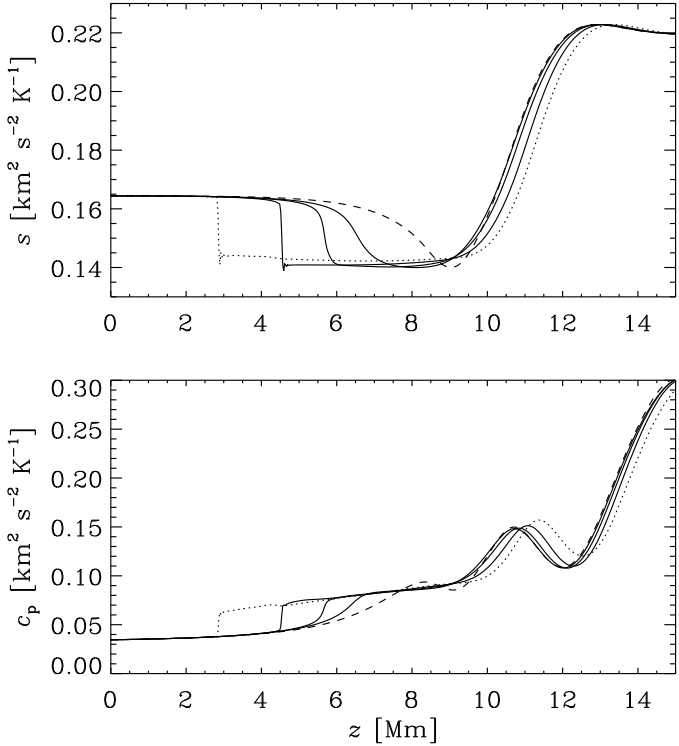}
\end{center}\caption[]{
Specific entropy and specific heat at constant pressure versus height,
from Run K3a.
}\label{pprofs_ss}\end{figure}

\subsection{$\Hminus$ opacity}

Finally, we compare with models using the $\Hminus$ opacity.
Again, we use here the implementation of \cite{HDNB06,HNSS07}, which
was found to yield reasonable agreement with realistic opacities.

\subsubsection{One-dimensional equilibrium models}

\begin{figure*}[t!]\begin{center}
\includegraphics[width=\textwidth]{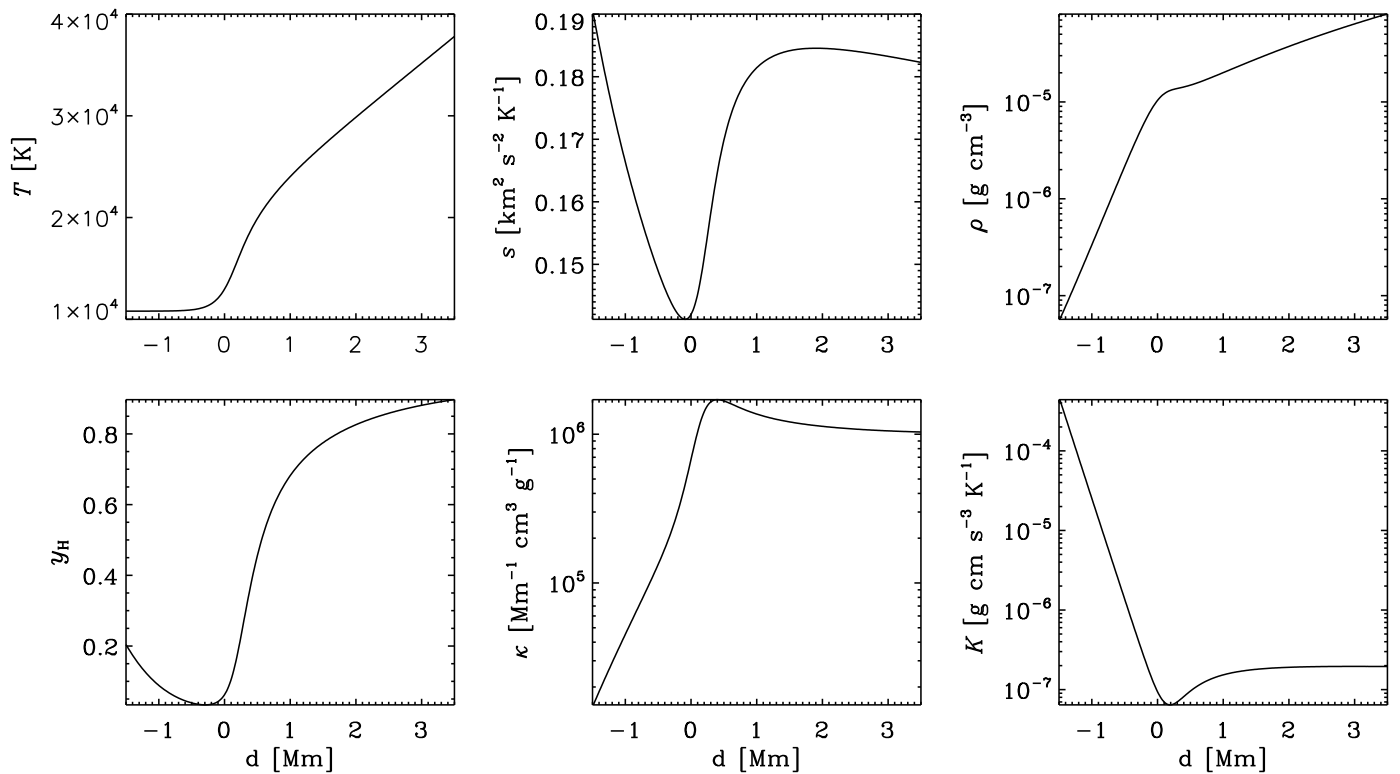}
\end{center}\caption[]{
Profiles of $T$, $s$, $\rho$, $\yH$, $\kappa$, and $K$ for
Run~D, with $\Hminus$ opacity. 
The abscissa gives the depth $d=z(\tau=1)-z$ in $\Mm$.
}\label{hmprofall}\end{figure*}

In \Fig{hmprofall}, we give 1D equilibrium solutions as functions of
depth focusing on the top $5\Mm$ (Run~D has a height of $9\Mm$,
where we have chosen $T_0=6\times10^4\K$).
The zero on the abscissa coincides with the $\tau=1$ surface
and depth $d=z(\tau=1)-z$.
We find a stably stratified lower part with an unstable part 
just beneath the $\tau=1$ surface.
The temperature decreases linearly from the bottom,
where $K$ is seen to be constant,
indicating the regime where the diffusion approximation applies,
similar to the other runs with Kramers opacity.
However, close to the $\tau=1$ surface
there is a short jump (decrease) in the temperature
by a factor of $\sim 2$, 
unlike the runs with Kramers opacity, where
the temperature profile simply turns 
from linearly decreasing to a constant value.
The temperature profile eventually settles to a constant 
for $z>z(\tau=1)$ or $d<0$.
This jump in the temperature profile resembles 
the profile in Fig.~1 and Fig.~14 in \cite{SN98}, 
where again the jump is by a factor $\sim 2$ in temperature.
It is attributed
to the extreme temperature sensitivity of the $\Hminus$ opacity.

For comparison, we include Run~E, for which we have chosen $T_0=10^5\K$
and a height of $20\Mm$.
The value of $z_{\tau=1}$ is then nearly $14\Mm$.
Now, however, there is an extended deeper layer which is stably stratified.

\subsubsection{Two-dimensional models}

\begin{figure}[h!]\begin{center}
\includegraphics[width=\columnwidth]{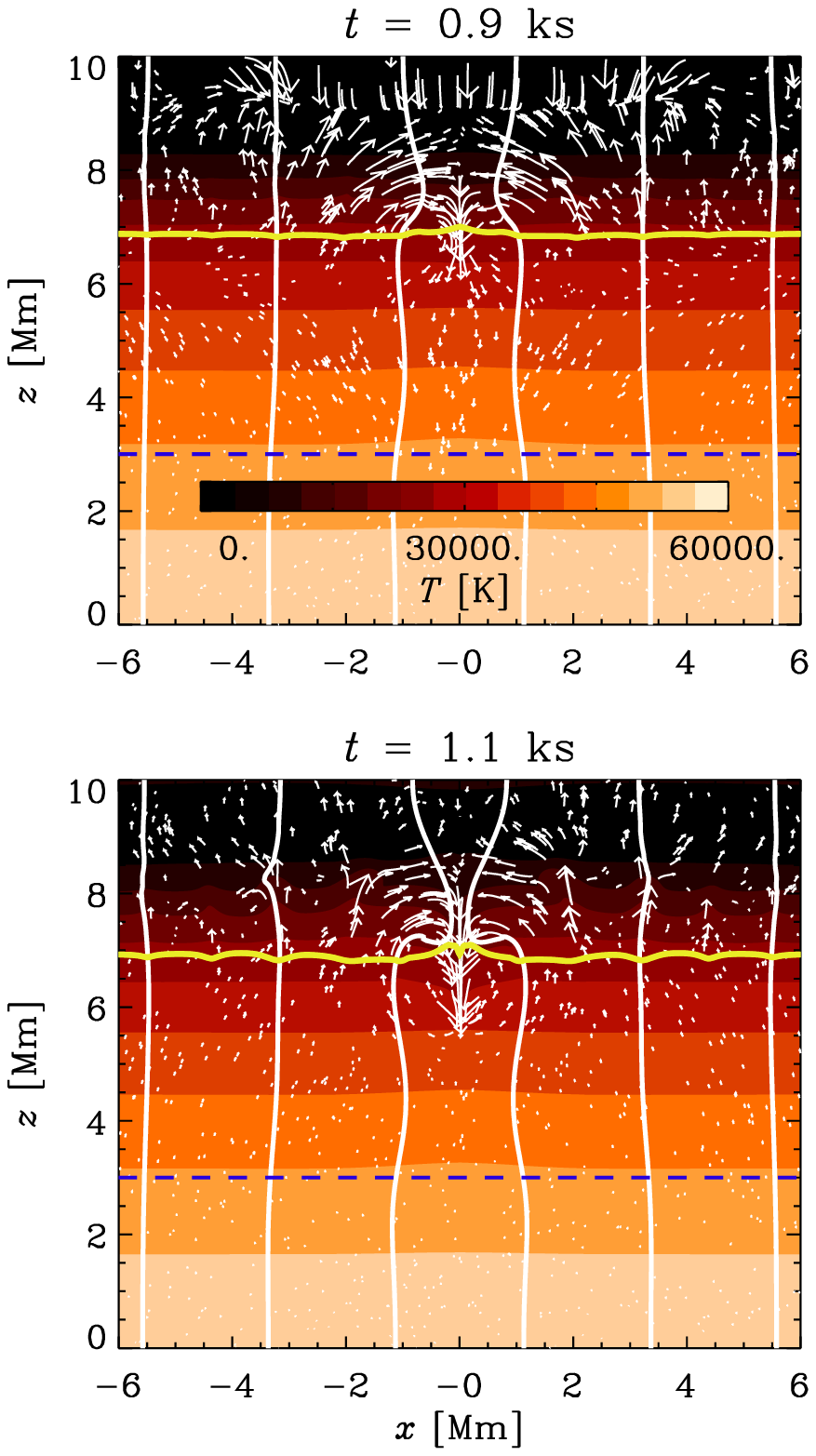}
\end{center}\caption[]{
Same as \Fig{pplane}, but for Run~H3 using the $\Hminus$ opacity
with a scaling coefficient of $10^{-6}$.
}\label{pplane_Hm}\end{figure}

In the 2D model with $\Hminus$ opacity, we chose
$\phi_0=-3\times10^{-3}\g\cm^{-3}\km^2\s^{-2}$
with $z_0=3\Mm$ for the height of the blob.
In \Fig{pplane_Hm}, we see that the flux concentrations 
form much above the blob location, 
close to the $\tau=1$ surface. This is again mainly
due to the negative gradient in entropy just below $\tau=1$ surface
as seen in \Fig{hmprofall}.
Furthermore, there is a very narrow dip
in the $\tau=1$ surface in the lower panel of \Fig{pplane_Hm} at $t=1.1\ks$,
but is flanked by two peaks, which is due to the return flows.

\begin{figure}[t!]\begin{center}
\includegraphics[width=\columnwidth]{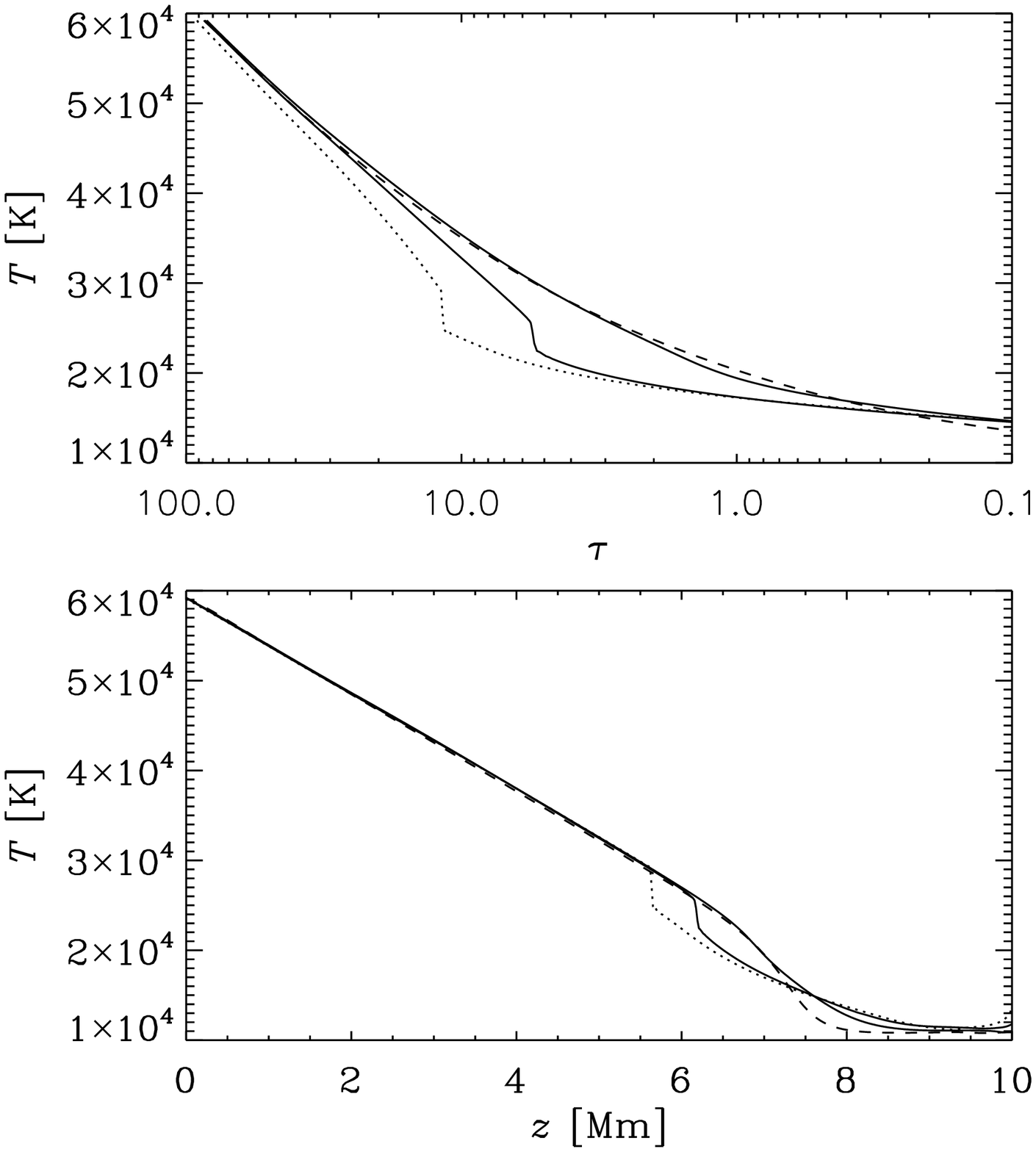}
\end{center}\caption[]{
Same as \Fig{pprofs_TT}, but for Run~H3 using the $\Hminus$ opacity
with a scaling coefficient of $10^{-6}$.
Note the increase in temperature at $z\approx8\Mm$.
}\label{hmtempevol}\end{figure}

Owing to the stable stratification of the lower part,
the resulting speeds are much lower than those in runs K3a and K3b.
As a consequence, the cooling in the temperature profile
due to the downflow of low entropy material, 
shown in \Fig{hmtempevol}, is decreased.
Compared to the case of Kramers opacity in \Fig{pprofs_TT}, 
most of the cooling here takes place to much lesser extent in depth.
This is further limited because 
the stratification soon becomes unstable towards larger values of $z$.

Comparing with the deeper model, where $T_0=10^5\K$ (Run~H10,
whose equilibrium model was Run~E), significant downflows can
only be obtained when we place the blob higher up ($z_0=10\Mm$) and
increase the forcing ($\phi_0=-3\times10^{-2}\g\cm^{-3}\km^2\s^{-2}$).
This is because of the more extended stably stratified deeper layer.
The maximum downflow speed is only $15\km\s^{-1}$.

\section{Conclusions}

The inclusion of partial ionization along with radiative transfer
forms an important step towards bridging the gap between idealized
models of magnetic flux concentrations and more realistic ones.
In this work, we have studied the effects of partial ionization firstly in 
1D hydrostatic models of the atmosphere in thermal equilibrium and
then in 2D hydraulic models of flux concentrations. In the radiative transfer 
module, we have used either Kramers opacity or $\Hminus$ opacity. 

Comparison of the final 1D equilibrium atmospheres
with and without partial ionization shows that, while the solutions
do not differ much in the optically thick part, they are significantly
different in the range $1<\tau<100$, especially with respect
to the specific entropy and density profiles.
An interesting feature is the narrow layer with a negative gradient
in specific entropy close to the $\tau=1$ surface, which is persistent
across different atmospheres
with either Kramers opacity (for any polytropic index; shown for $n$=
3.25, 1.5 and 1) or the $\Hminus$ opacity.
This minimum in the $s$ profile is directly
connected to the minimum in $\yH$ profile.
In fact from \Eqs{dsdh1}{dsdh2}, it is clear that the extrema in $s$
correspond to the extrema in $\yH$. This unstable layer
near $\tau=1$ is important since, in the 2D models, it causes the
flux concentrations to form right at the surface.

In 1D models with $\Hminus$ opacity, the $\tau<1$ part is
stably stratified as expected and here also a narrow unstable layer is seen
close to surface.
Due to the extreme sensitivity of the $\Hminus$ opacity
to temperature, there is a distinctive jump (by a factor $\sim2$)
in the temperature profile after a prolonged decrease.

In order to study the effect of partial ionization on hydraulic 
flux concentrations, the model we used employed an artificially imposed
source of negative pressure in the momentum equation.
This work has demonstrated that such a forcing function can lead to a
dramatic downflow that is channeled along vertical magnetic field lines.
A corresponding return flow is produced that converges in the upper
parts and draws vertical magnetic field lines together, which leads to
significant magnetic field amplification.
This strong amplification is connected with the high-speed descent of gas.
It is much faster than what is expected based on the artificially
applied pumping and it is in fact virtually independent of it.
Weaker forcing only leads to a delay in what later always tends to
develop into nearly free fall.
We do not expect such rapid descent speeds to occur in the Sun,
because there the gas is turbulent and will behave effectively in a much
more viscous and also more irregular fashion, where downdrafts break up
and change direction before they can reach significant speeds.

In the case of $\Hminus$ opacity, the flux concentrations are weaker
because the deeper parts are stably stratified.
Here again, the turbulence would have mixed the gas even before
triggering downflows, so the background stratification would be more
nearly adiabatic to begin with.
This can be seen clearly from realistic solar simulations of \cite{SN98};
see their Fig.~13. 

In models without partial ionization, flux concentrations form
just above the height where the forcing function is placed,
whereas in models including partial ionization, such flux concentrations
form at the surface (where $\tau=1$).
Here the specific entropy is unstably stratified and tends to drop by a
significant amount. Under the influence of downward suction,
this could still lead to significant descent speeds with a corresponding
return flow as a result of mass conservation.
The return flow, instead of closing
near the height where the forcing function is placed, closes at
the surface, from where the gas had earlier been pulled down.

It is surprising that the temperature reduction inside the downdrafts
is rather modest and to some extent compensated for by the supply of
hotter material from the converging return flow.
Thus, the magnetic structure is in our case largely confined by dynamic
rather than gas pressure.
Therefore the changes in the thermodynamic properties across the flux tube
are only moderate. As a consequence, the $\tau=1$ surface remains nearly flat.

In view of applications to sunspots, it would be important to consider the
effects of turbulent convection and its suppression by the magnetic field.
Such effects have been used in the models of \cite{KM00} that could explain
the self-amplification of magnetic flux by a mechanism somewhat reminiscent
of the negative effective magnetic pressure instability.
In our model, convection would of course develop automatically if we only
let the simulation run long enough, because the stratification is already
Schwarzschild unstable.
The degree to which the resulting convection contributes to the vertical
energy transport should increase with increasing opacity, but with the
rescaled opacities in our models it will be less than in the Sun.

Our findings also relate to the question of what drives convection in the
outer layers of the Sun.
Solving just the radiative equilibrium equations for the solar envelope would
result in a stable stratification, because the standard Kramers opacity
with $a=1$ and $b=-7/2$, corresponding to a stable polytrope with $n=3.25$.
Yet, those layers are unstable
mainly because of the continuous rain of low entropy material from the top.
Clearly, a more detailed investigation of this within the framework of the
present model would be needed, but this is well outside the scope
of the present paper.
Based on the results obtained in the present work, we can say that
the effects of partial ionization and resulting stratification are of crucial
importance for the production of strong magnetic flux amplifications just
near the visible surface.

\begin{acknowledgements}
PB thanks Nordita for support and warm hospitality
while work on this paper was being carried out.
She also acknowledges support from CSIR in India and use
of high performance facility at IUCAA, India.
This work was supported in part by the European Research Council under the
AstroDyn Research Project No.\ 227952, and
by the Swedish Research Council under the project grants
621-2011-5076 and 2012-5797.
We acknowledge the allocation of computing resources provided by the
Swedish National Allocations Committee at the Center for
Parallel Computers at the Royal Institute of Technology in
Stockholm and the National Supercomputer Centers in Link\"oping, the High
Performance Computing Center North in Ume\aa,
and the Nordic High Performance
Computing Center in Reykjavik.
\end{acknowledgements}

\appendix

\section{Thermodynamic functions}
\label{Thermo}

For completeness, we list here the relevant thermodynamic functions
as implemented by Tobias Heinemann into the {\sc Pencil Code}.
We have
\begin{equation}
c_p=\left({5\over2}+A_p B_p^2\right){{\cal R}\over\mu},\quad
c_v=\left({3\over2}+A_v B_v^2\right){{\cal R}\over\mu},
\end{equation}
as well as $\alpha=A_p/A_v$ and $\delta=1+A_p B_p$, where
\begin{equation}
\begin{array}{ll}
A_p={\displaystyle{\yH(1-\yH)\over(2-\yH)\xHe+2}},\quad&
B_p={\displaystyle{5\over2}+{\chi_{\rm H}\over\kB T}},\\ \\
A_v={\displaystyle{\yH(1-\yH)\over(2-\yH)(1+\yH+\xHe)}},\quad&
B_v={\displaystyle{3\over2}+{\chi_{\rm H}\over\kB T}}.
\label{AB}
\end{array}
\end{equation}

\section{Effect of partial ionization on entropy profile}
\label{ionssprof}

On differentiating the equation of state, $p={\cal R}T\rho/\mu$, we have,
\EQ
{\dd p}=\frac{{\cal R}T}{\mu}{\dd\rho} +\frac{\rho {\cal R}}{\mu}{\dd T}-\frac{\rho {\cal R}T}{\mu^2}{\dd\mu}.
\label{diffp}
\EN
Then we express $d\mu$ in terms of $d\yH$ using \Eq{mu},
\EQ
{\dd p}=\frac{\rho {\cal R}T}{\mu_0}\left[\left({\dd\ln\rho} +{\dd\ln T}\right)(1+\yH+\xHe)+{\dd\yH}\right].
\label{diffpfinal}
\EN
We substitute \Eq{diffpfinal} into the equation for first law of thermodynamics,
$T\dd s=\dd e + p\dd v$, where $v=\rho^{-1}$ is specific volume,
\begin{eqnarray}
{\dd s}=&&\frac{1}{T}\left[{\dd e} + {\dd(pv)} -\frac{1}{\rho}{\dd p}\right] \\
=&&\left(\frac{3{\cal R}}{2\mu_0}{\dd\ln T} -\frac{{\cal R}}{\mu_0}{\dd\ln\rho}\right) (1+\yH+\xHe) \nonumber \\ 
&&- \frac{{\cal R}}{\mu_0}{\dd\yH}\left[\frac{3}{2}+\frac{\chiH}{k_B T}\right],
\label{dseq}
\end{eqnarray}
where we have used $e=(3/2)\,{\cal {\cal R}}T/\mu+e_{\rm H}$.
Next, differentiate the Saha equation of ionization,
$\yH^2/(1-\yH)=R$, where
\EQ
R=({\rhoe/\rho})\left({T_{\rm H}/ T}\right)^{-3/2}
\exp\left(-{T_{\rm H}/ T}\right),
\EN
we have,
\EQ
{2\yH \over (1-\yH)} {\dd\yH} + {\yH^2 \over (1-\yH)^2} {\dd\yH } = \dd R
\label{dyeq}
\EN
and
\EQA
\label{dreq}
\dd R = {\rhoe/\rho}\left({T_{\rm H}/ T}\right)^{-3/2} \exp\left(-{T_{\rm H}/ T}\right) \times && \\ 
\left[-{\dd\ln\rho} -\frac{3}{2}{\dd\ln T}+ \frac{\chiH}{k_BT}{\dd\ln T}\right]. &&\nonumber
\ENA
After substituting \Eq{dreq} into \Eq{dyeq},
we obtain a relation between $\dd\yH$, $\dd\ln T$ and $\dd\ln\rho$,
\EQ
{(2-\yH)\over \yH(1-\yH)}{\dd\yH} = \left[-{\dd\ln\rho} -\left(\frac{3}{2}- \frac{\chiH}{k_BT}\right){\dd\ln T}\right].
\label{dyeqfinal}
\EN
Based on the behaviour of temperature profile we can have two cases, the optically
thick $\tau\gg1$ and the optically thin $\tau\ll1$.
In the case of $\tau \gg 1$, $\rho \propto T^n$ and hence, $\dd\ln\rho=n\,\dd\ln T$.
We use this relation in \Eq{dyeqfinal}, and have,
\EQ
{\dd\ln T}= -{1\over (n+1.5)}{(2-\yH)\over \yH(1-\yH)} \dd\yH
\label{dlnt}
\EN
In the optically thick part, T is large, thus $T_H(dT/T^2)$ is small
and can be neglected in \Eq{dyeqfinal}.
Again we use the relation $\dd\ln\rho=n\,\dd\ln T$ in \Eq{dseq} to eliminate $\dd\ln\rho$
and finally substitute \Eq{dlnt} into \Eq{dseq}, to obtain, 
\EQA
\dd s=\dd\yH{{\cal R}\over\mu_0}\left[{(n-1.5)\over(n+1.5)}{(2-\yH)(1+\yH+\xHe)\over\yH(1-\yH)}\right. && \nonumber \\
\left.- {3\over2} - {\chiH\over k_BT}\right].
\label{dsdh11}
\ENA
In the case of $\tau \ll 1$, we 
use the fact that $\dd\ln T=0$ as $T$ is nearly constant here and then obtain,
\EQ
\dd s=\dd\yH{{\cal R}\over\mu_0}\left[{(2-\yH)(1+\yH+\xHe)\over\yH(1-\yH)} - {3\over2} - {\chiH\over k_BT}\right]
\label{dsdh22}
\EN
Both \Eqs{dsdh11}{dsdh22} can be written in the following form using
expressions in \Eq{AB},
\EQA
\dd s=\dd\yH{{\cal R}\over\mu_0}\left[{1\over A_v}{(n-3/2)\over(n+3/2)}
- B_v\right], &&
\label{dsdh111}
\ENA
\EQ
\dd s=\dd\yH{{\cal R}\over\mu_0}\left[{1\over A_v}- B_v\right].
\label{dsdh222}
\EN
From \Eq{dsdh111} and \Eq{dsdh222}, its clear that the extrema in entropy
profile correspond to the extrema in $\yH$. 


\end{document}